\documentclass[10pt,journal,compsoc]{IEEEtran}

\ifCLASSOPTIONcompsoc
  \usepackage[nocompress]{cite}
\else
  \usepackage{cite}
\fi

\ifCLASSINFOpdf
  \usepackage[pdftex]{graphicx}
\else
  \usepackage[dvips]{graphicx}
\fi

\ifCLASSOPTIONcompsoc
  \usepackage[caption=false,font=footnotesize,labelfont=sf,textfont=sf]{subfig}
\else
  \usepackage[caption=false,font=footnotesize]{subfig}
\fi

\begin{document}

\title{Scale Stain: Multi-Resolution Feature Enhancement in Pathology Visualization}

\author{Jesper~Molin,
	    Anna~Bod\'{e}n,
        Darren~Treanor,
        Morten~Fjeld,
        and~Claes~Lundstr\"{o}m
\IEEEcompsocitemizethanks{\IEEEcompsocthanksitem Jesper Molin is at Chalmers University of Technology, CMIV at Link\"{o}ping University, and at Sectra AB, Sweden.\protect\\
E-mail: jesper.molin+tvcg@gmail.com
\IEEEcompsocthanksitem Anna~Bod\'{e}n is at the Department of Clinical Pathology and Department of Clinical and Experimental Medicine, Link\"{o}ping University.
\IEEEcompsocthanksitem Darren~Treanor is at Leeds Teaching Hosiptal NHS Trust, United Kingdom and at CMIV, Link\"{o}ping University.
\IEEEcompsocthanksitem Morten~Fjeld is at Chalmers University of Technology.
\IEEEcompsocthanksitem Claes~Lundstr\"{o}m is at CMIV, Link\"{o}ping University and at Sectra AB.}
}


\IEEEtitleabstractindextext{%
\begin{abstract}
Digital whole-slide images of pathological tissue samples have recently become feasible for use within routine diagnostic practice. These gigapixel sized images enable pathologists to perform reviews using computer workstations instead of microscopes. Existing workstations visualize scanned images by providing a zoomable image space that reproduces the capabilities of the microscope. This paper presents a novel visualization approach that enables filtering of the scale-space according to color preference. The visualization method reveals diagnostically important patterns that are otherwise not visible. The paper demonstrates how this approach has been implemented into a fully functional prototype that lets the user navigate the visualization parameter space in real time. The prototype was evaluated for two common clinical tasks with eight pathologists in a within-subjects study. The data reveal that task efficiency increased by 15\% using the prototype, with maintained accuracy. By analyzing behavioral strategies, it was possible to conclude that efficiency gain was caused by a reduction of the panning needed to perform systematic search of the images. The prototype system was well received by the pathologists who did not detect any risks that would hinder use in clinical routine.
\end{abstract}

}

\maketitle

\IEEEdisplaynontitleabstractindextext
\IEEEpeerreviewmaketitle

\IEEEraisesectionheading{\section{Introduction}\label{sec:introduction}}


\IEEEPARstart{I}{mage-generating} technologies are essential tools within modern medicine. Images generated by different modalities, most notably Computer Tomography or Magnetic Resonance Imagining scanners, are analyzed by medical doctors through visual inspection. For diseases where analysis at the cellular level is needed, pathology imaging is instead used. Pathology imaging is an invasive technique including removal, processing and visualization of tissue samples from the body.

Up to date, most tissue samples have been analyzed by pathologists who review them using a light microscope. Recently it has instead become possible to create digital images of the tissue sample using a new modality called Whole-Slide Imaging scanners. These scanners consist of a movable microscope that traverses the specimen at high magnification and stitches together gigapixel-sized images with sub-micrometer resolution. A modern scanner can typically scan a high quality image in minutes, which has contributed to why this modality has gained traction and is currently being implemented for clinical use to replace diagnostic microscope review \cite{Pantanowitz2015} in several laboratories worldwide. The scanners generate image pyramid files with the scanned image in multiple scales, which make it possible to quickly retrieve and view the image at any magnification and location. The fastest viewing systems can retrieve the images fast enough that it takes the same amount of time to perform the diagnostic review as with the microscope \cite{Randell2014b}. This new practice has spun off a new field of study: Digital Pathology, which investigates the new possibilities that are made possible by being able to generate these digital microscopic images. New uses of the technology include teaching, remote viewing, as well as automated image analysis to assist the diagnostic review \cite{Pantanowitz2015}.

Within other medical imaging domains it is common practice to visualize the imaging data in different ways depending on the task. For example, to find lung nodules or to review circulation of contrast medium in Computed Tomography images, a specific projection can be applied that extract the most important parts of the large image volume. While these types of projections have since long been explored by the scientific visualization community, they have so far not been explored for pathology images.

For volume data, important patterns are often hidden by the fact that outer layers such as the skin occlude the inner organs. For pathology images that are essentially two-dimensional, important patterns are instead hidden in the vast size of the image -- image features that are too small to be visible at low magnification. Conventionally, both with the microscope and digital pathology workstations, pathologists explore these patterns by physical or virtual panning and zooming.

\begin{figure*}[ht!]
\includegraphics[width=\textwidth]{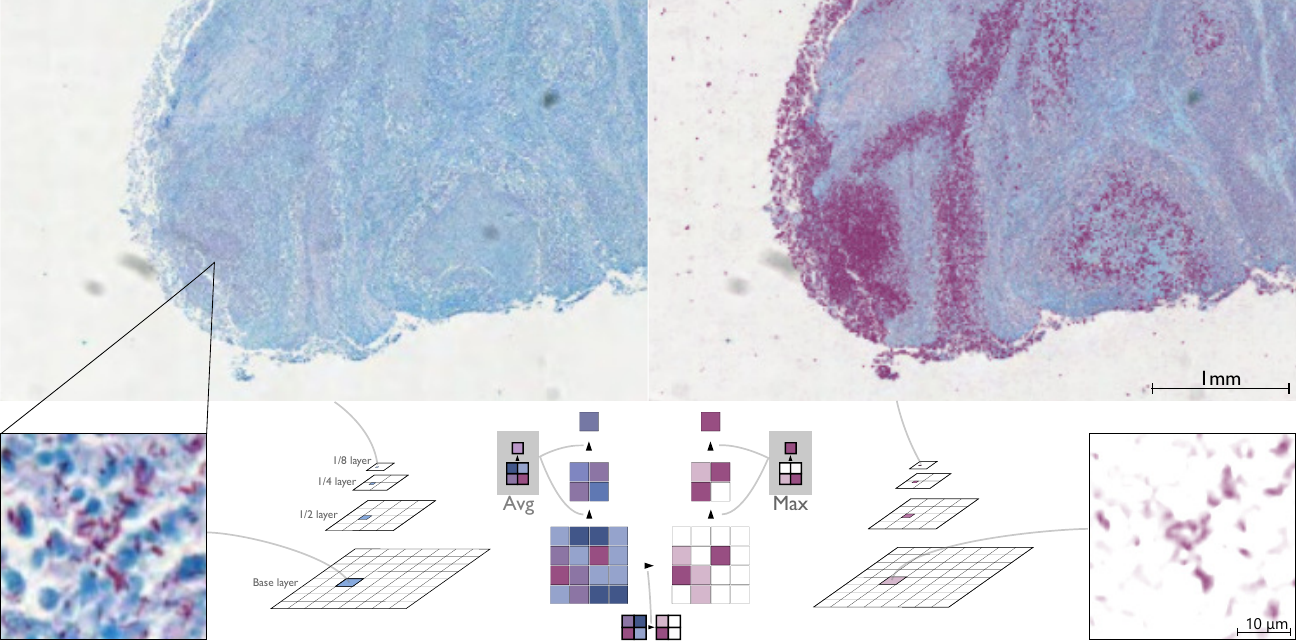}
\caption{The figure to the left shows a microscopic image of a lung biopsy with Tuberculosis bacteria stained in purple. In order to make these visible in lower magnification, in the right figure the Scale Stain visualization pipeline is used, which is based on color deconvolution and max-value subsampling. In section 4, a detailed explanation of how the pipeline works is provided.}
\label{fig:teaserFigure}
\end{figure*}

In this paper, the idea of alternative projections is brought the world of pathology. To exemplify this idea, a novel visualization pipeline, a prototype system and the results from a user study with the prototype are presented. The visualization pipeline, named Scale Stain, makes it possible to extract image features of a particular color that are otherwise not visible at at low magnification. Hence, the novel pipeline enables pathologists to perform diagnostic tasks that were previously not possible at low magnification. The key principles of this novel pipeline are illustrated in Figure \ref{fig:teaserFigure}.

The paper is organized as follows: First, a description is presented of how pathology images are generated to understand how the existing visualization pipeline works, this is followed by a short summary of what pathologists look at in the pathology images. Second, related work within medical image visualization, information visualization and digital pathology is presented. Third, the Scale Stain visualization pipeline  is described together with its implementation details. A user study is then presented with eight participating pathologists, where the efficiency, accuracy and usage strategies are evaluated when performing two typical diagnostic tasks. Finally, we discuss the results of the user study and possible use cases and limitations.

\section{Background}
To instrument the visualization system for pathology images, we started by analyzing how current microscopic images are generated and reviewed. As a background for the methods presented later, we will here provide an overview of these clinical processes. Only the typical pipeline is covered, leaving out side-tracks and special cases.

\subsection{The pathology imaging pipeline}
The purpose of pathology imaging is to produce microscopic images from tissue. The most common types of tissue samples subjected to pathological examination are biopsies and surgical specimens. Biopsies are small tissue samples that are removed, e.g. from the breast or from skin, by a hollow needle or excised from the tissue. Surgical specimens are tissue that is removed under surgical procedures, such as when a part of a lung is removed because of detected tumor.

These tissue samples are then processed in different chemical compounds to preserve and fix the tissue, and are then embedded in paraffin in order to create stable blocks. From the paraffin blocks, representative thin sections of around 1-5 micrometers are sliced with a microtome and placed on 1x3 inch glass slides.

\begin{figure}
\includegraphics[width=\columnwidth]{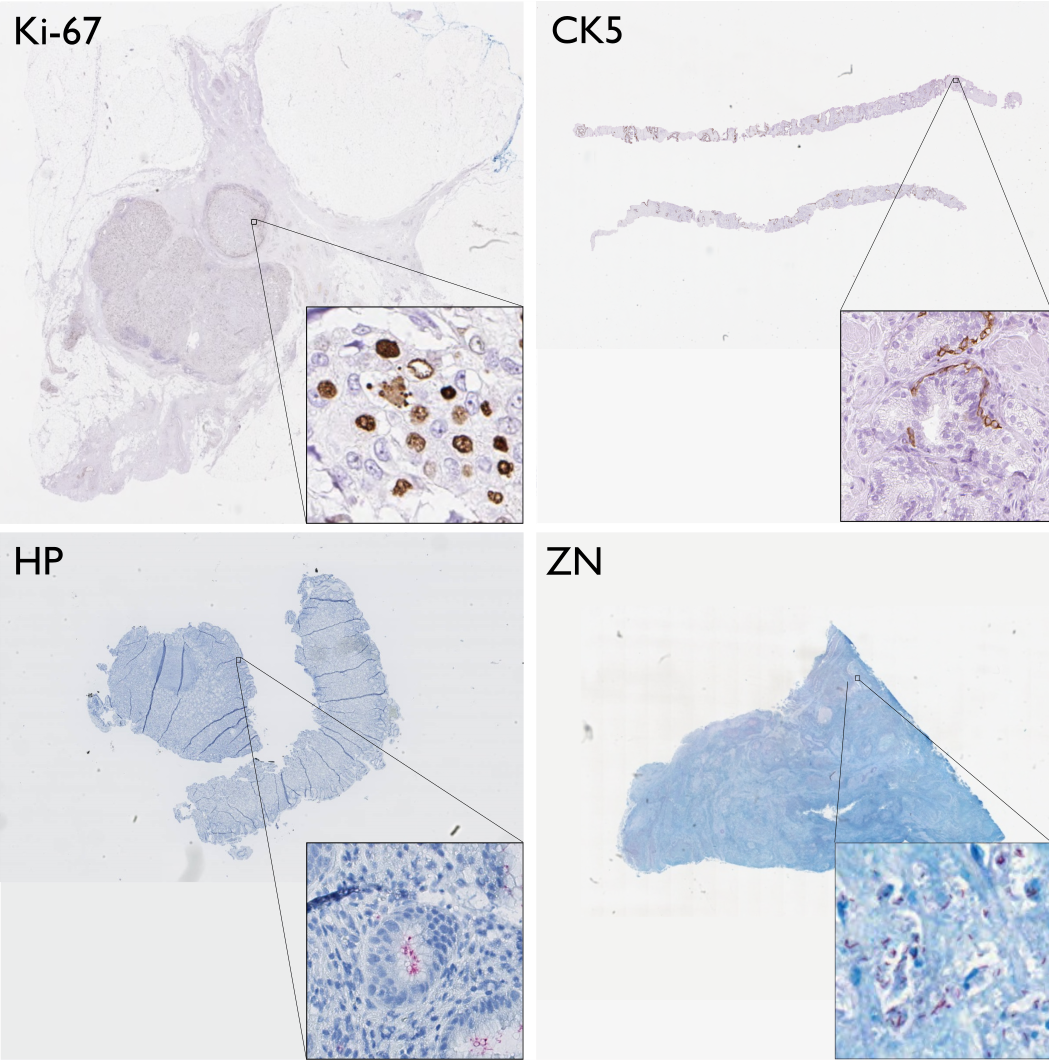}
\caption{Examples of stains with small colored objects that is not visible at low magnification. Top-left: Ki-67 stains a protein associated with proliferation in the cell nuclei in a sample of breast tumor. Top-right: CK-5 stains basal cells normally surrounding benign glands in a prostate tissue biopsy. Bottom-left: HP stains Helicobacter Pylori bacteria in a gastric biopsy. Bottom-right: ZN stains tuberculosis bacteria in a lung biopsy.}
\label{fig:ihcExample}
\end{figure}

At this stage of the process, the tissue is almost completely transparent. To increase the visibility of important structures, different chemical stains are used that give color to the tissue. Hematoxylin \& Eosin (H\&E), the most common stain (around 80\% of the slides at our lab), stains acid structures blue and basic structures pink. Other stains are more specific, including “special” histochemical stains or immunohistochemistry stains. In many applications these stains only stain the tissue sparsely, as can be seen in Figure \ref{fig:ihcExample}.

After the glass slides have been stained, they are organized in cases on trays and delivered to the pathologists who review them using a microscope. In a digital workflow, the slides are instead scanned in a whole-slide imaging scanner. A high-end scanner can be loaded with racks of glass slides and then uses a robotic microscope stage to capture image patches across the slide at high magnification and stitch them together to a large digital image. Most scanners are able to capture images of the tissue at 400 times magnification that are sampled at around 0.25 microns per pixel. By convention, 400 times magnification is denoted 40x, since the eye-piece magnifies the image 10 times in a conventional microscope. For the standard sized tissue piece of 15x15mm, the resulting size of the digital image is 3.6 gigapixels. In order to be able display and navigate these digital images at microscope-like speed, the images are divided into subsampled tiled pyramids. To display an image view at a particular location and magnification, tiles are requested from the pyramid and stitched together to fill up the size of the image display. Using dyadic pyramids increases the file size by 33\%, since every magnification level is 1/4 of the size of the previous one. This overhead is needed to ensure quick panning and zooming in the gigapixel-image, since it would take too much time to perform the subsampling operation on the fly. Overall this pipeline enables pathologists to perform diagnostic review with a computer workstation instead of a microscope.

\subsection{The diagnostic review}
The diagnostic review is performed both macroscopically before the tissue is processed and microscopically using the glass slides. The majority of the microscopic review is performed on H\&E slides, where pathologists detect visual patterns and combine that with their knowledge about different diseases. The output of the diagnostic review is a written report that is sent back to the referring physician.
This microscopic review is performed by navigating around in the images, identifying important structures and findings, note absence of other findings, interpreting what has been identified, and forming and confirming different hypotheses \cite{Crowley2003}.

The typical microscopic review starts by inspecting the tissue at low magnification to locate areas of interest, followed by navigating the slide in medium to high magnification to confirm findings. This search is then mixed with occasional zoom actions when needed \cite{Ruddle2016,molin2014b}.

Besides H\&E stained slides, it is also possible to use special stains for specific diagnostic tasks, like estimating cell proliferation rate or discriminating between possible differential diagnoses. In Figure \ref{fig:ihcExample}, four different stains are presented: Ki-67, CK5, HP, and ZN.

\textit{Ki-67} is a protein that is found in the cell nuclei during proliferation. By staining for this protein, the proliferation of a tumor sample can be established by counting the number of brown-stained tumor cells. The recommended diagnostic protocol for breast cancer in Sweden states that the percentage for tumor cells that are stained should be counted and reported in the hotspot of the tumor, meaning the area with 200 cells that contains the largest number of positive cells.

\textit{CK5} is a subtype of keratin that, for example, is found some types of basal cells. In the depicted prostate biopsy, the staining is used to detect whether the possibly malignant glands are surrounded with basal cells or not. Lack of basal cells indicates that the glands may be cancerous.

\textit{HP} or Helicobacter staining is used on gastric biopsies to stain the Helicobacter Pylori bacteria, which is associated with Gastritis or gastric ulcers. The pathologists review the whole slide to determine whether bacteria are present or absent.

\textit{ZN} or Ziehl-Neelsen staining is used to stain for acid-fast bacteria, and is commonly used in the diagnosic review of Tuberculosis. Similarly to the HP stain, the pathologists determine absence or presence. However, the biopsies are usually larger and the organisms more sparsely dispersed, which make this task even more laborious.

Besides these four examples, a typical pathology laboratory has access to hundreds of stains that are used for different purposes. This complexity is however manageable since different stains share common characteristics. The same background staining is used (often Hematoxylin, which is blue), and a primary staining with a specific target that can be, for example, brown or red. The target stain, can stain the nuclei, the cytoplasm, the membrane or combinations of these. It is also possible to target different cell types.

This study will focus on the review of these stains, which can be both time consuming and complicated. The stains are commonly not visible in low magnification. This means that in order to detect the staining, the pathologist needs to zoom in to high magnification and pan through the whole slide.

The pathologists also need to look out for artifacts, which commonly occur due to variations in the handling of the specimen caused by the surgeon or in the laboratory process. The pathologists therefore make sure that a positively stained tissue makes sense in the context of the diagnostic review and double check with staining-independent morphological features in the tissue. For example, to conclude presence of bacteria, both the rod-shaped form of the microbe and positive staining needs to be detected.

\section{Related work}
The previous work most relevant for our proposed method comes from three different categories: Volume visualization techniques for 3D medical datasets, multi-scale systems within information visualization, and a smaller body of semi-automation and color calibration within pathology.

\subsection{Visualization in medical imaging}
It is common practice to review medical images by adjusting different visualization parameters. Brightness and contrast adjustments are common for flat x-ray images, and transfer functions support the review of volume data.

Basic volume data visualization techniques include Maximum Intensity Projections (MIP) and Direct Volume Rendering (DVR). The MIP work by preserving the voxel value with the highest value through casted rays, which create flat projections that highlight the contrast medium.  In DVR, different voxel values are assigned  to different colors and projected to create 3D-renderings, which highlight different features useful for the diagnostic review \cite{Fishman2006}. These visualization methods are useful because they provide a simple way for radiologists to reduce a large image dataset to something that fit on the display for the specific diagnostic task.

Several advances of these methods have been explored. Viola et al. \cite{Viola2005} presented a method to weight the visibility of different image objects based on a predefined importance function, forcing specific features to become visible. Bruckner, Gr\"{o}ller \cite{Bruckner2009} formulated a method combining the benefits from MIP and DVR. Both methods counteract the occlusion of small image objects that can otherwise be hard to distinguish with normal DVR.

Another way to raise the salience of important visual features is to modify their apparent size. Wang et al. \cite{Wang2011} enlarged image regions based on a user-selected transfer function. Correa and Ma \cite{Correa2008,Correa2009} experimented with local size and occlusion of objects as an additional parameter to the transfer function to increase different discrimination possibilities between image features. Our method also extracts important details from a large dataset, but brings this capability to the domain of large two-dimensional images.

\subsection{Multi-scale visualization}
Even though pathology images are naturalistic depictions of the tissue, working with them share common points with visualization of large datasets, especially when multi-scale visualization methods are used to depict the data. Generalized Fisheye views is the idea that for a large dataset you can apply a degree-of-interest function to all data-points that weights the visibility of each point \cite{Furnas1986}. Another important concept is Semantic zooming  \cite{Perlin1993}, which denotes the capability of changing what information is visible depending on the zoom-level, for example, in a modern digital map where the names of cities are shown at low magnification, and street-names and buildings are prioritized at high magnification. Shneiderman \cite{Schneiderman1996} summarizes the typical tasks performed in large information spaces in the Visual Information Seeking Mantra: “Overview first, zoom and filter, then details-on-demand”. This type of functionality has been implemented in systems a large variety of datasets. Perhaps most similar to pathology images are matrix visualizations (e.g. \cite{Elmqvist2008,Behrisch2014}) or large geographically distributed data (e.g. \cite{Goodwin2016}). A summary of multi-scale systems is provided by Elmqvist, Fekete \cite{Elmqvist2010a}, who modeled the visualization of multi-scale representations, including both different ways to interact with the data and different ways to aggregate the data into representative views.

These multi-scale visualization techniques are important to understand and adopt for gigapixel-sized images, which also operate in a multi-scale visualization space. Important concepts can be reused, but need to be adapted in order to work for image data.

\subsection{Pathology visualization}
Earlier research has focused on different automatic approaches and color calibration methods. Automatic methods have been developed for many of the described stain applications. For one of the typical applications, Ki-67 hotspot selection, several automatic methods to select hotspots exist \cite{Swiderska2015,Niazi2016,Roullier2011}. These methods can be helpful, but the accuracy is generally not sufficient without double checking the result. The algorithms yield a more accurate result in that they generate higher percentages for the hotspot selection task \cite{Swiderska2015}, but it is still assumed that pathologists are better at detecting false positives \cite{Niazi2016}. Using more advanced laboratory procedures, like using multiple parallel stains can increase the accuracy of automatic counts \cite{Stalhammar2016}, but this does not remove the need for appropriate visualization tools in order to understand the underlying data and to increase the efficiency of the review.

Previous visualization work for pathology imaging has focused on improving the speed of viewing these gigapixel images \cite{Jeong2010}, or dealing with 3D-stacks of microscopic images \cite{Hadwiger2012}. For the day to day needs of a pathology lab, commercially available scanners can capture, stitch and organize digital slides into tiled pyramidal images in minutes \cite{scannercomp2012}, and digital viewers that can closely reproduce the experience of using a microscope \cite{Randell2014b}. However, it is important to keep viewing latency down, where the bottlenecks typically are slow hard drives \cite{Jeong2010} and slow network response times.

Work on color reproducibility within pathology visualization involves several studies dealing with normalizing the variation caused by using different scanners, different batches of stains, and other process related issues in the staining laboratory. These approaches \cite{Khan2014} often use color deconvolution matrices to separate out the different staining contributions to the pixel value \cite{ruifrok2001quantification}. Bejnordi et al. \cite{EhteshamiBejnordi2016} is recommended for a comprehensive list of staining normalization methods.

Image enhancement methods have been proposed by Landini, Perryer \cite{Landini2009} to improve the image for color-blind people, and by Kather et al. \cite{Kather2015}, who created a method to that extends the hue range for a specific staining combination thus improving the perceptual contrast. These methods improve the visual acuity in the highest magnification but are not efficient for enhancements at low magnification. Contrary to these approaches, this paper focuses on image enhancements at low magnification by including visual information from higher magnification levels of the image within the enhancement method.

\section{The Scale Stain pipeline}
Our visualization problem can be seen in two ways. Either as a low magnification image that can be enhanced by using information from high magnification levels, or as a large gigapixel-sized image that is reduced to a smaller image. In the rest of the paper, the latter view is adopted. This means that the problem can be seen as finding a mapping function that maps a gigapixel-sized image to a representative image that fits onto a normal computer display, reducing the amount of image information with three orders of magnitude while keeping the information that is relevant for a particular task.

When designing this mapping function, a prototype was first built and evaluated in a feasibility study \cite{Molin2014a}. Working closely together with pathologists, three requirements were iteratively derived:

\begin{itemize}
\item[R1.] The reduction function should be 100\% sensitive.
\item[R2.] The staining density rank should be preserved.
\item[R3.] The appearance of the representative image should be connected to the appearance of the full magnification image.
\end{itemize}

The reason to require \textit{R1} and \textit{R2} can be understood directly from the task descriptions for the different stains. \textit{R1} is needed in order to support the microbial detection for the HP and ZN staining. It is not possible to conclude that stained images features are absent unless the mapping function preserves the staining with 100\% accuracy. The second requirement \textit{(R2)} is needed to support the hotspot selection task for Ki-67 stained slides. To perform this task, the pathologists need to compare the density of the staining between different regions of the image slide. This means that it is a sufficient requirement if it is possible to determine whether one area is denser than another, but not with how much.

The third requirement \textit{(R3)} is important in order to give the pathologists a full sense of control. This control is mediated by connecting the appearance of the high and low magnification view, by two problem solving mechanisms. First, the appearance matching makes it easier to detect false positives when zooming in, since the connection between low and high magnification is made intuitively. Second, when zooming out after inspecting a particular feature in high magnification, it is possible to extrapolate the information gained to similar areas in low magnification, even though these areas are not visited explicitly. To make this connection possible it is important that the mapping between the image and its low-level representation is predictive and easily understood.

All requirements are not needed for all applications, but by implementing a mapping function that fulfills all of them the final visualization system becomes more general.

\subsection{Mapping function design}
There exist multiple mapping functions that fulfill all requirements. Both usability factors and the ease of the technical implementation were considered when designing the mapping function adopted in this paper. The mapping function is described in the form of processing pipeline, which consist of three steps: color deconvolution at the base level, extracting the important information with max-value subsampling, and finally blending the extracted information with the original image.

\subsubsection{Color deconvolution}
At a high magnification where the staining is clearly visible, the amount of staining is extracted in each pixel. The extraction uses the color deconvolution method by Ruifrok, Johnston \cite{ruifrok2001quantification}. The method estimates the amount of color molecules present in the specimen by applying Lambert-Beers law and measures the correlation between the RGB color in each pixel to a pre-defined reference color. Since different stainings can be quite similar, glass slides are always labeled with what staining has been used so the approximate reference color is almost always known. By applying color deconvolution to the original image, a map of how much staining there is in each pixel in the high magnification is created. This is an analogue to the importance map concept presented by Viola et al. \cite{Viola2005} -- the pixels with most staining are the pixels that are most important to visualize.

\subsubsection{Max-value subsampling}
The importance map from the previous step was extracted at high magnification, so it now has to be reduced to the size of computer display. In the standard pipeline this operation is performed using Gaussian subsampling, which acts as an average subsampling function: it approximately takes four neighboring pixels and maps the average color onto a single pixel in order to halve the size of the image. Because many tissue slides are only sparsely stained with the target color, the colors are washed out. Instead max-value subsampling is applied, which is defined in this paper as taking four neighboring pixels and transfer the pixel with the maximum value onto a single pixel. For an image that is scanned at 40x, this reduction is repeated 5 times in order to fit the gigapixel-image onto a typical display. It is equivalent to dividing the high-magnification image into 32x32 neighboring areas and extract the maximum value directly. The advantage of doing this in five steps instead, is that the intermediate levels can be used to create a smooth transition between the high and low-magnification image when the user zooms in, thus fulfilling \textit{R3}.
Max-value subsampling is non-linear, but it is possible to show for an ideal case that it is both 100\% sensitive and that it retains the rank order between different areas, which is necessary to fulfill \textit{R1} and \textit{R2}.

\subsubsection{Importance map blending}
In the final step, the max-value importance map (the output from the previous step) is multiplied with the target color in RGB-space and blended on top of the original image at low magnification using alpha blending.

\begin{figure}
\includegraphics[width=\columnwidth]{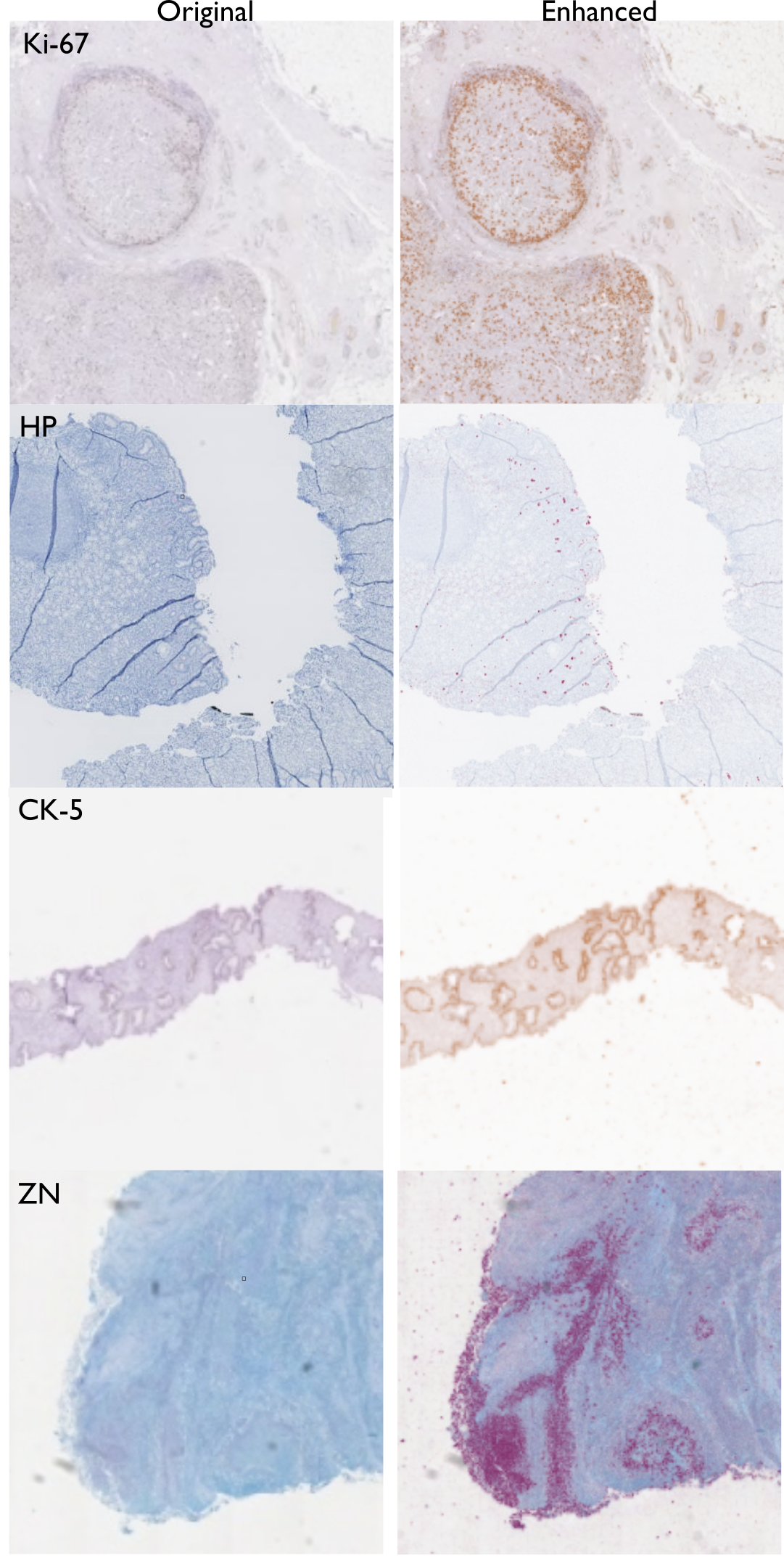}
\caption{Example of the enhanced images compared to the original of four different stainings. The patches are taken from the same slides as the slides in Figure \ref{fig:ihcExample}. To the left in each pair is the unmodified image, and to the left, an enhanced image with optimized visualization settings.}
\label{fig:scalestainExamples}
\end{figure}

The effect of these three steps creates a low magnification representation of the gigapixel-image where the sparse staining component is clearly visible. Example results can be seen in Figure 
 \ref{fig:scalestainExamples}, where the processing has been applied to the images presented in Figure \ref{fig:ihcExample}, starting from a magnification level appropriate for each image.

At low magnification, it is useful to blend the importance on top of the original image, but at higher magnifications that utility decreases. Therefore, the prototype reduces the  amount of blending as the user zooms in, and at the base level the original image is always displayed. Since the importance map was created by adding multiple steps of max-value subsampling, it is also possible to render a smooth transition between low and high magnification level. These two mechanisms together, create an effect of \textit{zooming into the original image}, which provides a convenient way to verify the result of the mapping function.

\subsection{Visualization parameter space design}
The color deconvolution step in the visualization pipeline can start from any of the pre-generated magnification levels included in the image file created by the scanner. This makes it possible to use the start magnification as a sensitivity parameter. If the process starts at the maximum magnification of an image, the stained pixels are guaranteed to be visible in the representative image. However, if the staining density is too high, the representative image will become saturated, and it will be harder to compare the intensity between different regions. Instead, if the max-value subsampling starts at medium magnification, the staining has already been averaged out a couple of times, which washes out the staining.

\begin{figure}
\includegraphics[width=\columnwidth]{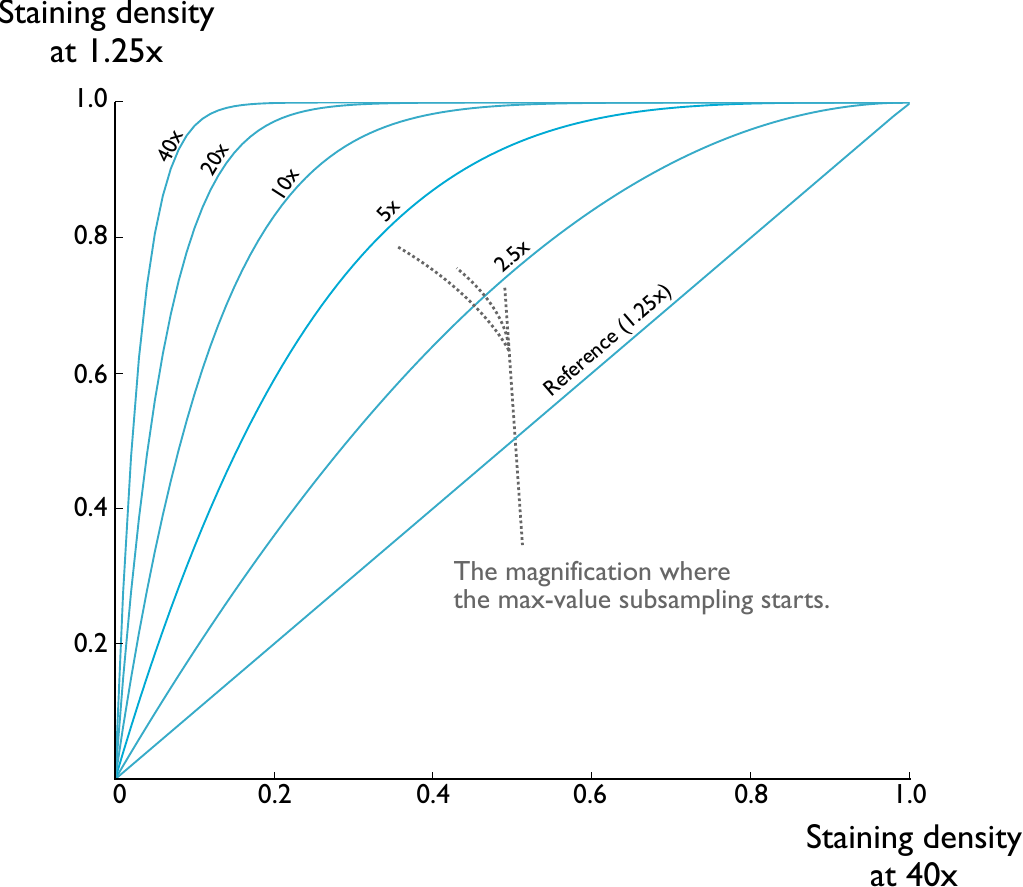}
\caption{Max-value subsampling have a non-linear contrast enhancing effect for cases where the staining density ratio in the original image is low. Note that the curves are expected pixel density means, since the max-value process is stochastic, i.e. each curve also have a variance that has been omitted.}
\label{fig:maxvalueDynRange}
\end{figure}

To further understand the sensitivity effect of max-value subsampling, it can be compared with the analogous max-pooling concept commonly used within machine learning for convolutional neural nets. If the pixel value distribution of the importance map pixels is approximated as independent and binomially distributed, the behavior of this mapping function can be plotted as in Figure \ref{fig:maxvalueDynRange}, using the derivation of the expected mean of the output of the max-pooling function by Boureau et al. \cite{Boureau2010}. The curves show that performing max-value subsampling has a non-linear contrast enhancing effect for sparsely stained images. The curves are also monotonously increasing, which means the rank order of the density is maintained. For most pathology slides, it is appropriate to estimate the pixel value distribution as binomial, but pixels next to each other are hardly independent in high magnification. Thus, the figure show the theoretical behavior in a similar situation.

Another advantage with varying the sensitivity this way, is that it is possible to explain the sensitivity dimension to the user: A sensitivity at the 5x level means that, if the staining is visible at 5x, it will always be visible in the representative image. Pathologists know from experience what image features are visible at what magnification, by using this sensitivity measure, that experience can be mapped onto the new visualization pipeline.

Another visualization parameter that is proposed is how the max-value importance map is blended with the original image. Three different blending modes are important: \textit{show only the original image}, \textit{show importance map on top of the original image}, or \textit{show only the importance map}. These blending modes can then be used as anchor points, and by linearly interpolating between these points it is possible to create a \textit{blending factor} dimension.

\begin{figure}
\includegraphics[width=\columnwidth]{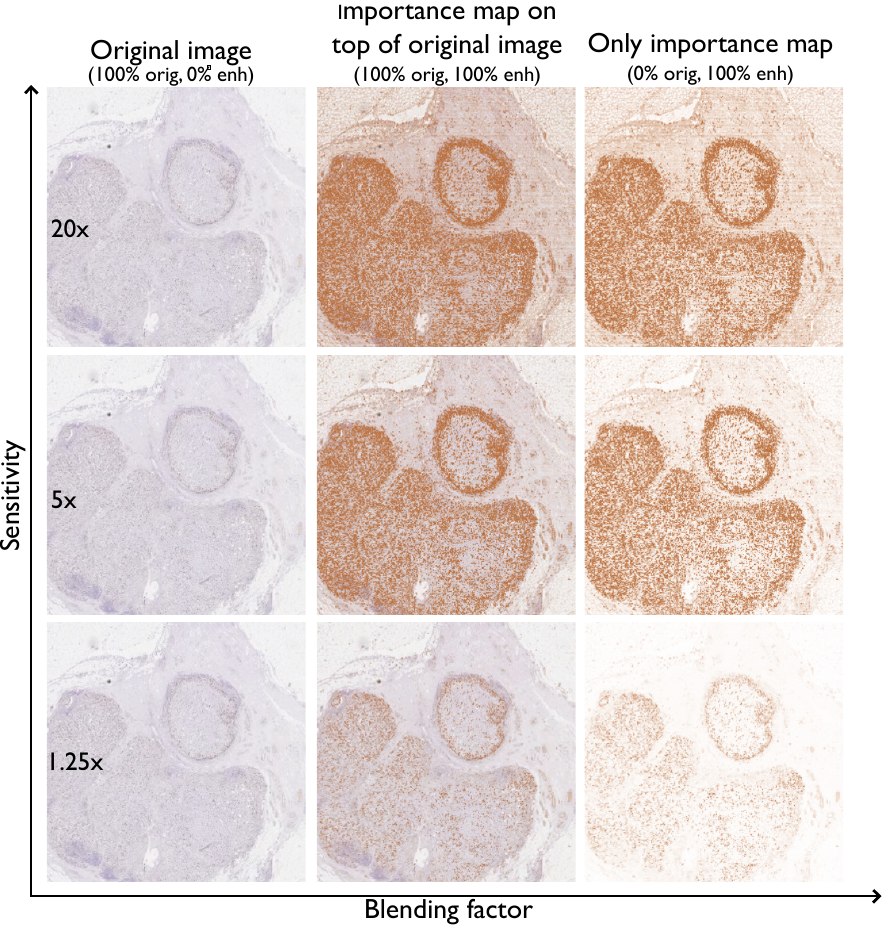}
\caption{From left to right, different blending factors (orig=\textit{percentage of original image shown}, enh=\textit{percentage of the enhanced importance map that rendered}). From top to bottom the sensitivity dimension is shown as the magnification where the color deconvolution was performed.}
\label{fig:parameterSpace}
\end{figure}

Together, the sensitivity and blending factor yield a two-dimensional parameter space that the pathologist can use to decide how to visualize the tissue. In Figure \ref{fig:parameterSpace}, nine different visualization modes from this parameter space are displayed for a Ki-67 slide.

\begin{figure}
\includegraphics[width=\columnwidth]{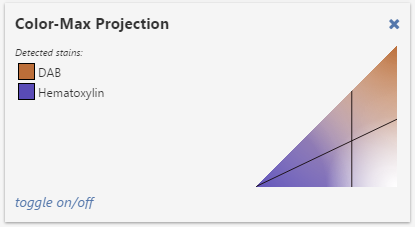}
\caption{To navigate the two-dimensional parameter space, a triangle shaped picker tool was designed. The sensitivity dimension goes from top to bottom, and the blending factor goes from left to right.}
\label{fig:parameterSpacePicker}
\end{figure}

The two dimensions are loosely coupled. If the blending factor is zero, the sensitivity factor has no effect on the visualization. To reflect this in the user interface, the parameter space picker was created in the form of a triangle, shown in Figure \ref{fig:parameterSpacePicker}. The x-axis controls the blending factor, and the y-axis controls the sensitivity. The sensitivity dimension is compressed to the left to reflect that varying the sensitivity has little effect when the blending factor is low. The dimensions are not explained explicitly, instead the parameter space is visualized by making a gradient between, background staining color, the target staining color and white as a way to intuitively represent the effect that the different parameter settings have on the image display. This intuition can be seen by comparing Figure \ref{fig:parameterSpace} and Figure \ref{fig:parameterSpacePicker}, imagining a compression of the sensitivity for low blending factors.

\subsection{Implementation details}
The process of deconvolving the gigapixel-sized image, applying max-value subsampling and blending it with the original image cannot be performed in real time using modern hardware. Instead, a part of the computations needs to be pre-processed.

\begin{figure}
\includegraphics[width=\columnwidth]{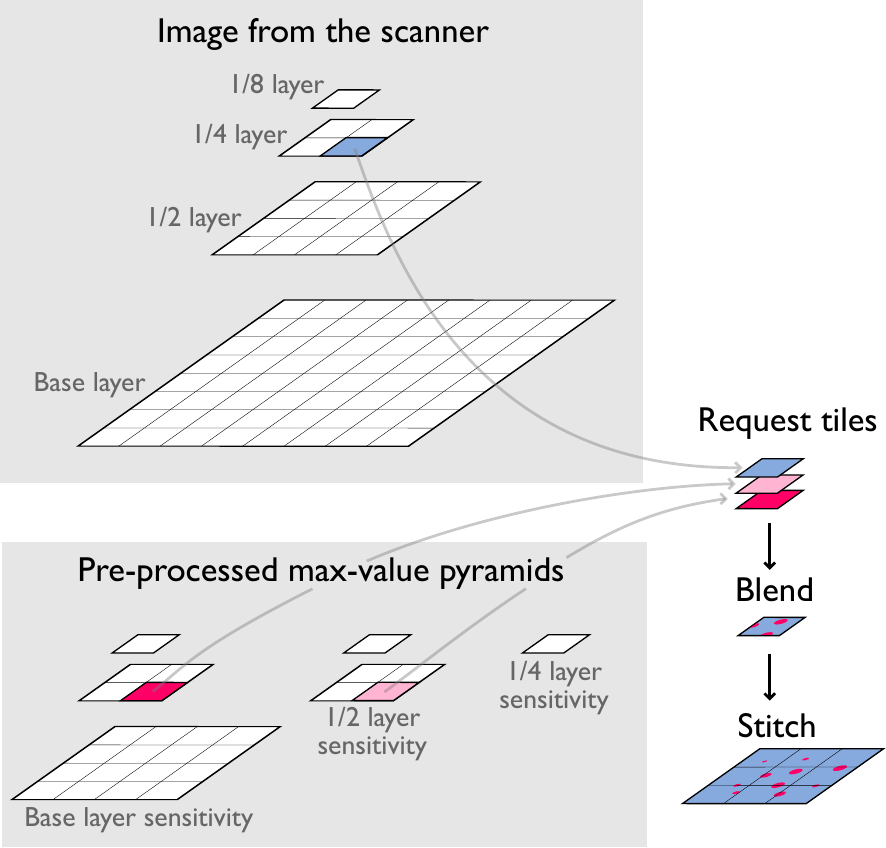}
\caption{In order to allow real-time exploration of the parameter space, part of the pipeline needs to be pre-processed. All color deconvolution and max-value computations are pre-processed whereas the blending between computed tiles and the original images is performed on the fly.}
\label{fig:preprocessing}
\end{figure}

The way that the visualization pipeline is set up, it is quite natural to pre-process all the color deconvolution and max-value subsampling operations, and store the result in separate pyramids beside the pyramid file of the original scanner image. Then when a certain image view needs to be viewed, the tiles from both the pyramid of the original image and the pre-processed pyramids are requested from disk, and blended together on the graphics card in the client software, is in Figure \ref{fig:preprocessing}.

The pre-processing increases the amount of needed storage. The number of additional tiles can in fact be derived exactly. Using double geometrical sums, both for the series of sensitivity pyramids and the series of levels for the individual pyramids, the number of tiles for all sensitivity pyramids are 50\% of the original image pyramid. The high magnification tiles require much storage but are quick to process. Therefore, a convenient way to save storage space is to drop the pre-processing of the tiles with the highest magnification in each sensitivity pyramid, decreasing the overhead to 12.5\%. The number of tiles does not directly translate to storage cost because of how well the different types of tiles are compressed. The original image tiles in our test set was compressed with JPEG at quality level 70, and we opted to use 8-bit greyscale PNG to compress the max-value tiles. For the Helicobacter stained images, which were the most sparse, the pre-processed pyramid was on average 36\% the size of the original pyramid. For the Ki-67, the same number was 43\%. Including the optimization described above, this means that overall storage overhead of the resulting visualization system lies at around 10\%.

The pre-processing computations can be highly parallelized. Our implementation uses all the cores on the processor and performs the color deconvolution on the GPU. This results in pre-processing rate of 1 Gpixels per 79 seconds on a Dell XPS 15 laptop (Intel Core i7-3632QM, 8Gb RAM, Nvidia GeForce GT 640M, LITEONIT LCT-512M3S SSD drive). The computation time for slides measured on SVS-files (a common file format) consisted of 49\% file I/O, 33\% color deconvolution, 10\% max-value subsampling and 6\% other computations). The processing time is very sensitive to fast tile access, which means that less efficient file formats can increase the overall processing time significantly. 

Overall, this technical design allows deployment of this visualization system into existing production systems, by adding a computational node that generates the max-value pyramids directly after images are scanned, at around 10\% storage overhead for images that would benefit for this type of visualization.

\section{User evaluation}
To evaluate the performance of the visualization system in a real context, a user study with professional users was performed. The aim of the study was to investigate the impact the system had on task efficiency and accuracy for two types of staining and to gauge potential users' perception of the system.

\subsection{Participants}
Eight pathologists, four specialists and four trainees were recruited by email request from the pathology labs at Link\"{o}ping University Hospital and Karlstad Central Hospital in Sweden. Five participants used digital images for review at least every week, while the remaining three pathologists did not.

\subsection{Apparatus}
\begin{figure}
\includegraphics[width=\columnwidth]{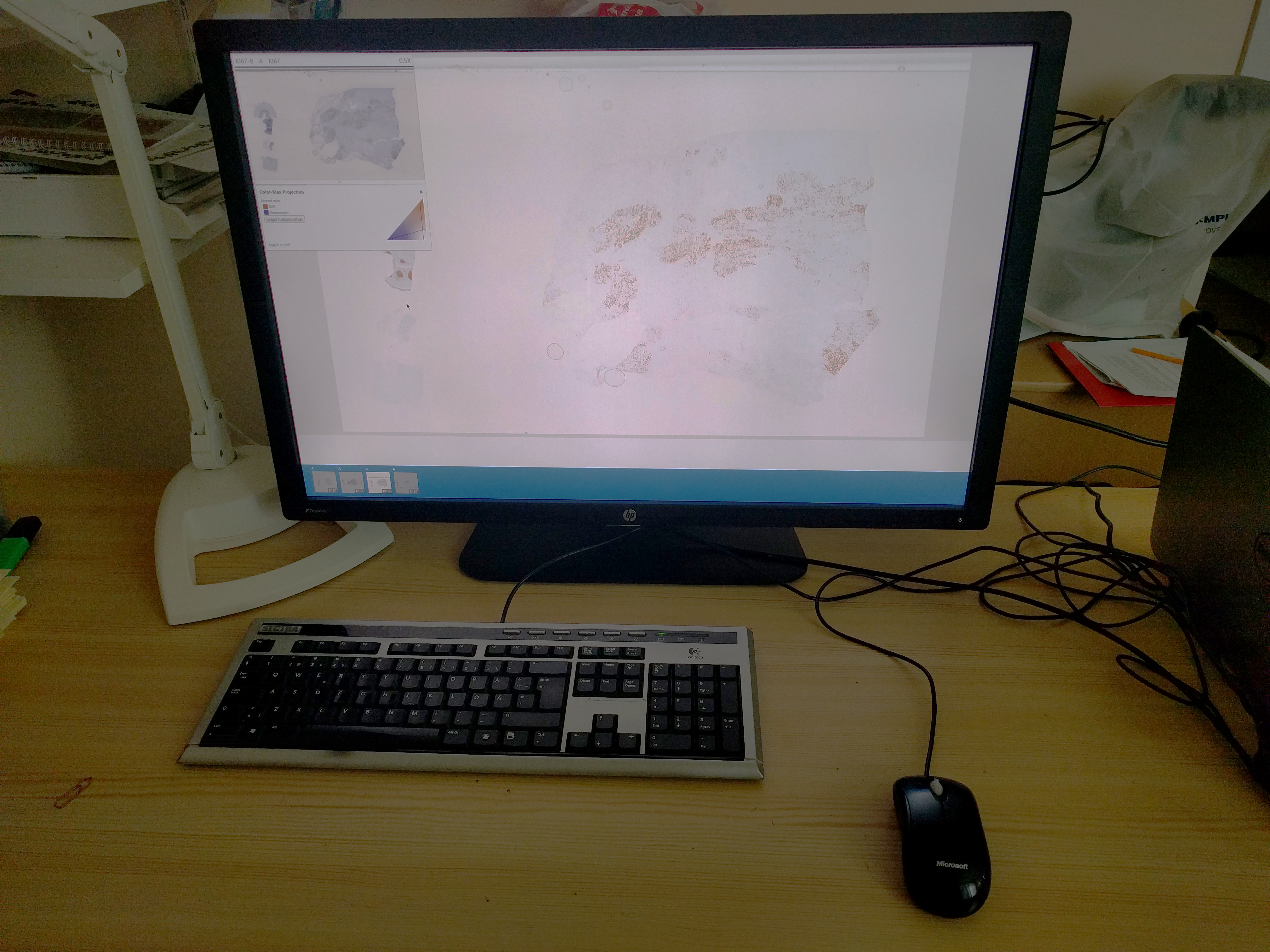}
\caption{Photo of the study setup on one of the two sites, where the user study was performed.}
\label{fig:experimentalSetup}
\end{figure}

The prototype software were running on version 18.1 of IDS7/px (Sectra AB) on a XPS 15 Laptop (Dell Inc). To navigate the slides, a standard two-button mouse with a clickable scroll wheel was used. The slides were reviewed on a 30 inch, 4 megapixel display (HP Z30i), with the color temperature set at 6500K, 100\% Brightness and 80\% contrast. A photo of the setup is given in Figure \ref{fig:experimentalSetup}.

\subsection{Cases and task}
The prototype was evaluated for two of the applications described in the background section: Helicobacter Pylori detection and Ki-67 hotspot selection. Two case sets were selected for this purpose.

From the 2015 production of scanned digital slides at the Link\"{o}ping University Hospital, 885 gastric biopsies were diagnosed, out of these, 85 were Helicobacter positive. From this positive set, 10 cases were randomly selected disregarding 17 cases because the staining was visible at low magnification, and 2 cases for being out of focus. This set was matched with 10 randomly selected negative Helicobacter cases.

The second set of cases were also taken from the 2015 production and consisted of 30 randomly selected breast tumors from a total of 180 cases that had been stained with Ki-67. No cases were excluded since the task could be performed manually on all the selected slides.

\subsection{Procedure}
The participants were first welcomed and the overall structure of the experiment was described to them. They then signed a written consent form together with a form that asked about their level of experience as a pathologist.

The prototype functionality was then demonstrated, what effect the visualization technique had on the display of the slides, how to control the visualization parameters, and what exploration strategies that might be useful. The participants were then allowed to freely use the tool until they clearly stated that they felt comfortable using the tool by performing the tasks on a few test slides. The participants were then asked to perform the task as if the trial had started, to ensure that they had found a strategy they felt comfortable with before the trial started. The participants were informed that the duration for each task was recorded but that it was more important to make a correct decision than performing the review quickly.

The participants started the Helicobacter task using the first set of 10 slides, with or without using the visualization tool. Then, the second set of 10 slides were reviewed with the opposite technique. The participants were instructed to navigate around in the slides to determine whether each slide was positive or negative for the Helicobacter bacteria and state their response out loud. The order of technique and slide set was fully counter-balanced to ensure that no slide was reviewed twice by the same participant and to avoid order effects. After the trials with the Helicobacter tasks, the participants were given the opportunity to take a short break before continuing with the next task.

In the Ki-67 task, the participants were asked to navigate around in each tumor slide and select a hotspot. The response were recorded by placing a fixed width circle around the selected area approximately containing 200 cells. The participants performed the task on 15 slides with and without the tool. The use of technique and slide set was fully counter balanced. The task completion time was measured from when the slide was opened, until the response were given for both tasks.

After the last task, a semi-structured debriefing interview was held. In the interview, usability and diagnostic safety were discussed, as well as reconnecting to different exploration strategies that were observed by the experimenter during the trials. All the user interaction with the system in terms of navigation in the slides and modifications of the parameter settings were automatically logged, and the training session, the trial and the final debriefing were audio recorded.

\subsection{Study design}
The experiment was a within-subjects design with technique (with two levels: reference, Scale Stain) as independent variable and task completion time and error rate as dependent variables. The task completion time from both tasks were combined into 2x2 experiment with task as an independent variable, whereas the error rate was treated as a one-way experiment for each task due to the fact different types of error rates were recorded for each task. For the Helicobacter task, it was considered an error if the negative/positive response disagreed with the expert controlled consensus. For the Ki-67 task, the error was measured as the absolute percentage difference from the hotspot selection of tumor cells with the highest percentage. Both the technique and the case variables were counter balanced using a Latin square. In total \textit{(10 helicobacter slides + 15 Ki-67 slides) x 2 techniques x 8 participants = 400 trials} were recorded, but one trial had to be removed due to a logging error.

Due to large number of repetitions per participant linear mixed effects analysis was used to test whether the technique had a significant effect on task completion time and accuracy. The statistical analysis was performed using \textit{R} and \textit{lme4}, and used study design informed maximal random-effect structures were used as recommended by Barr et al. \cite{Barr2013} for hypothesis testing. The task completion time model therefore used a random intercept for slides and by-participant and by-task random slopes. The Ki-67 accuracy model used a random intercept for slides and a by-participant random slope. The Helicobacter accuracy model used logistic mixed effects analysis and used random intercepts for slides and participant. Visual inspection of residual plots revealed a slight exponential effect for larger task completion times, but log correction did not affect the estimated P-values. No other obvious deviations from homoscedasticity or normality were observed. P-values were obtained by likelihood ratio tests of the full model against the model without the technique variable.

\section{Results}
\begin{figure}
\includegraphics[width=\columnwidth]{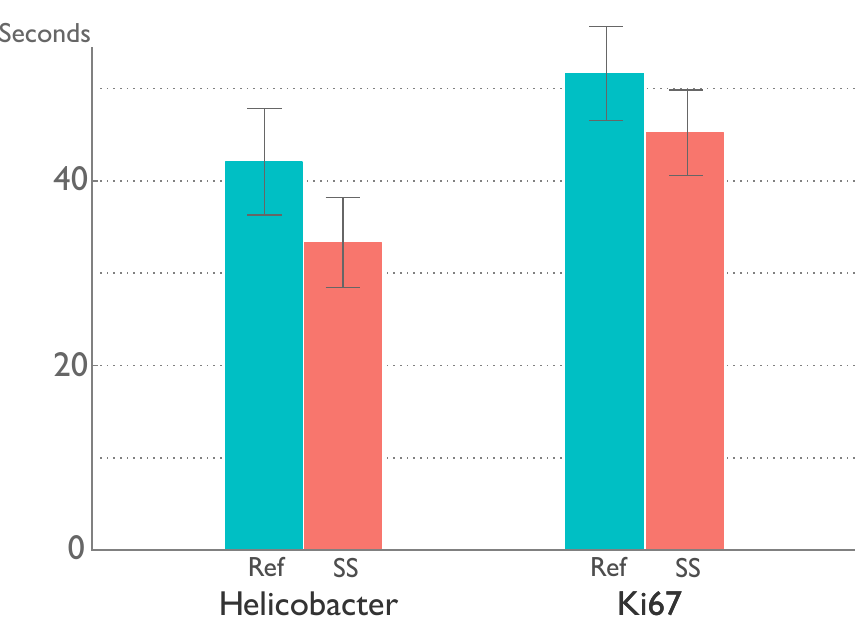}
\caption{Average task duration and standard error without (Ref) and with the Scale Stain technique (SS), as estimated by the fitted linear mixed effects model for both tasks and techniques.}
\label{fig:durations}
\end{figure}

\begin{figure}
\includegraphics[width=\columnwidth]{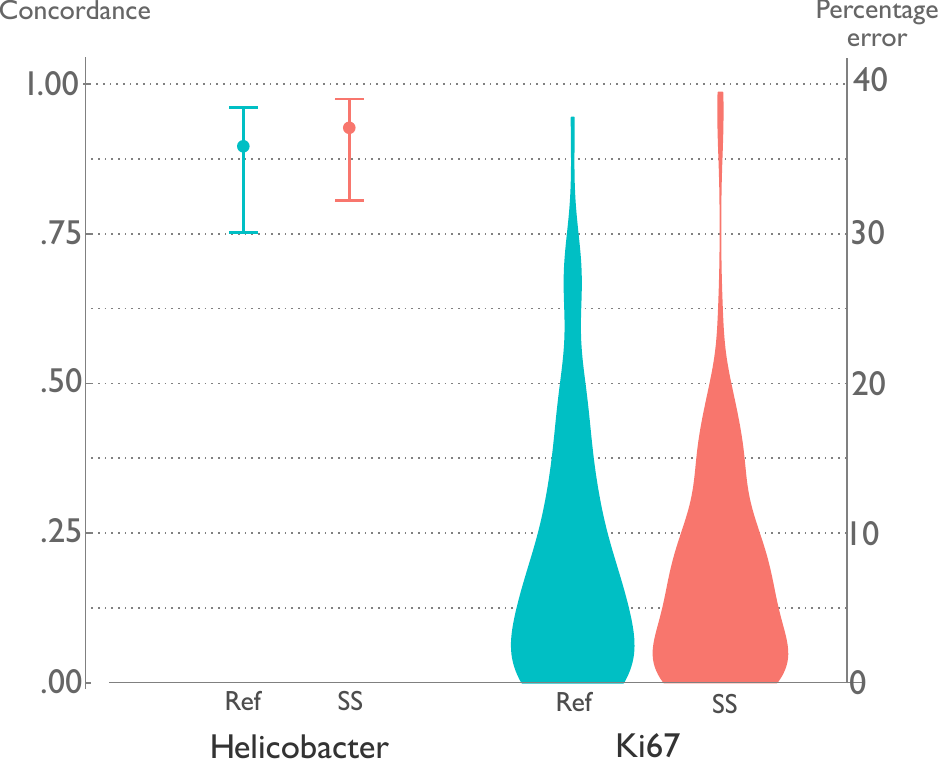}
\caption{The spread of the accuracy for both tasks without (Ref) and with the Scale Stain technique (SS). To the left, the accuracy of the Helicobacter task is presented as the 95\% confidence interval for both technique as estimated by the logistic regression mixed effects model. To the right, a violin plot (similar to a histogram) shows the absolute counting error distribution for both techniques.}
\label{fig:accuracy}
\end{figure}

The use of the visualization tool resulted in a shorter task completion time $({\chi}^{2}(1)=4.79, p=0.029)$. The shortening amounted to $7.4\pm2.7s$, corresponding to a $15\%$ shortening of the average time. The average task completion time was $44.1\pm25.9s$. There was no significant effect on the accuracy for either the Ki-67 task $({\chi}^{2}(1)=0.68, p=0.32)$ or the Helicobactor task $({\chi}^{2}(1)=0.56, p=0.45)$. The median error rate for the Ki-67 task was $5.9\%$ (IQR: $9.6\%$). For the Helicobacter task, the overall concordance rate was $88.1\%$, two participants had a perfect concordance with the consensus and the least concordant participant had $65\%$. The durations for each task and technique are given in Figure \ref{fig:durations} and the accuracies given in Figure \ref{fig:accuracy}.

\subsection{Exploration strategy}
\begin{figure*}
\subfloat[Reference condition]{
\includegraphics[width=\columnwidth]{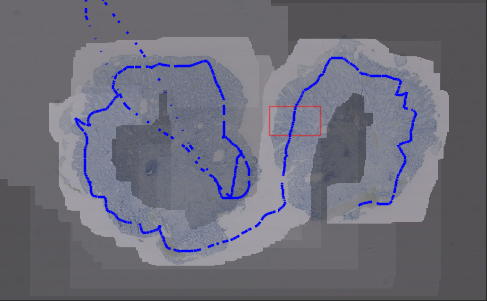}}
\subfloat[Scale Stain condition]{
\includegraphics[width=\columnwidth]{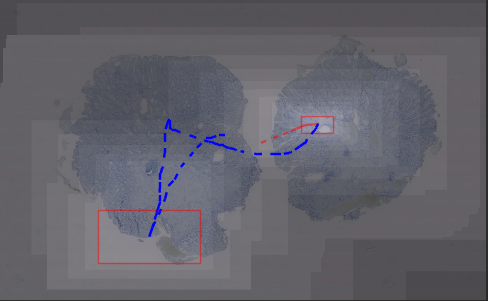}}
\caption{The typical strategy becomes apparant in a negative Helicobacter slide. Blue signify panning, and red zooming, a red square means that a user has zoomed in on location and then directly zoomed out again, and the slide overlay becomes brighter when a particular area has viewed in a higher magnification. In (a), a participant in the reference condition explores the edge of the biopsy in medium magnification where bacteria are usually found. In (b), a participant uses the tool to find regions of interest, and then only zooms in on those before deciding the biopsy is negative.}
\label{fig:navigationTracks}
\end{figure*}

The strategy used to solve the tasks using the Scale Stain technique was quite different from the strategy used in the reference condition. With the Scale Stain technique, most participants started reviewing the slide in low magnification, changing the visualization parameters if needed. This was followed by exploring interesting regions in the slide by zooming in and out and inspect them in high magnification, until a final decision could be made. In the reference condition, the main strategy consisted of zooming in to medium magnification and then scan the whole slide to search for bacteria or the hotspot. An example of this difference is given in Figure \ref{fig:navigationTracks}.

The technique used had the largest effect on the exploration strategy, but each participant also had their own idiosyncratic behavior. Five of the eight participants used the parameter space triangle to initially search the contrast at low magnification and to tweak the visualization settings for each individual slide. \textit{P2}, used the same strategy but the initial search period was much longer than for the others. Another strategy was used by one participant \textit{(P5)}, who stuck with the same setting for most slides but explored the parameter space extensively when needed. The remaining participant \textit{(P1)} used the same setting for almost all the slides.

\begin{figure*}
\includegraphics[width=\textwidth]{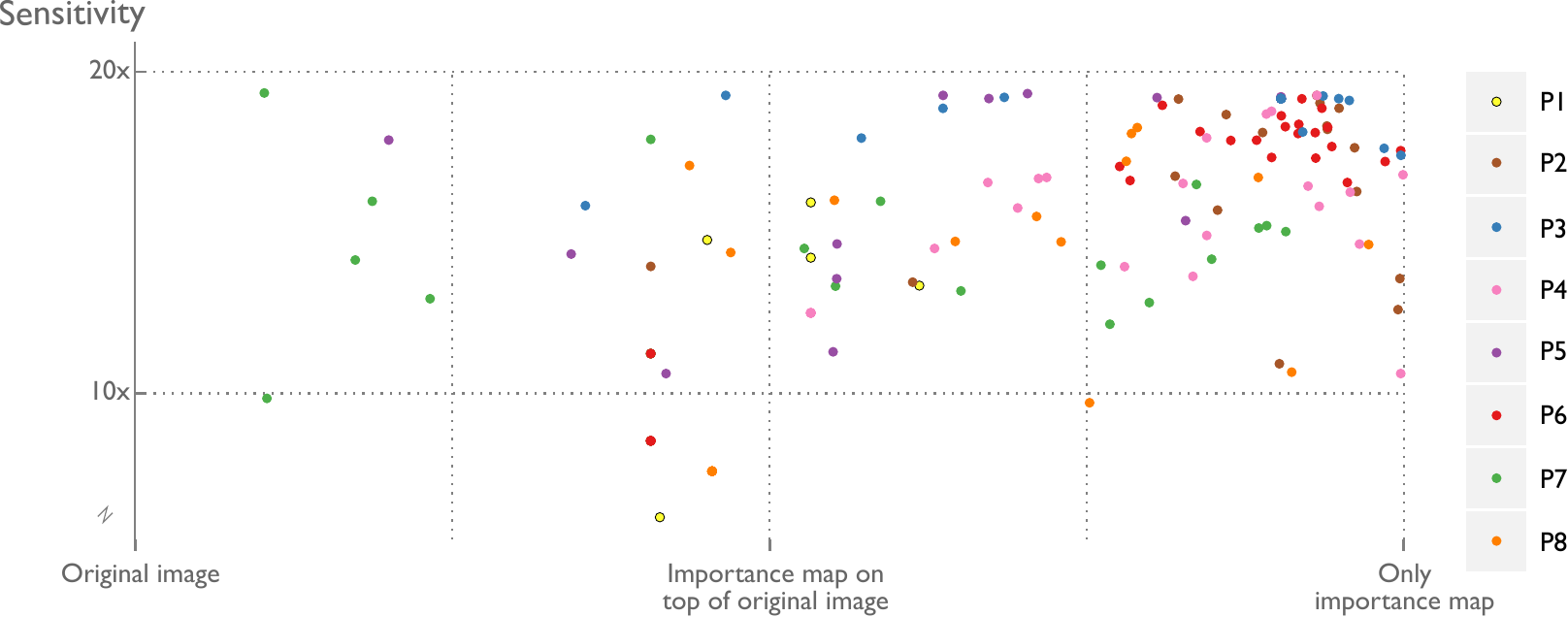}
\caption{The figure shows the final setting for each slide and participant after the parameter setting exploration phase in the case review. Most data points are in the area with high sensitivity showing that the enhancement effect was in fact used. The figure also shows the difference in preference between participants, e.g. compare P6 and P7. }
\label{fig:finalParameterSettings}
\end{figure*}

Even though the participants used a large part of the parameter space during exploration, the final parameter setting for each slide used the top half of the possible sensitivity levels. There was almost no difference between the final parameter setting between the Helicobacter and the Ki-67 task, instead each participant used their own set of final parameter settings,  which is depicted in Figure \ref{fig:finalParameterSettings}.

\begin{figure}
\includegraphics[width=\columnwidth]{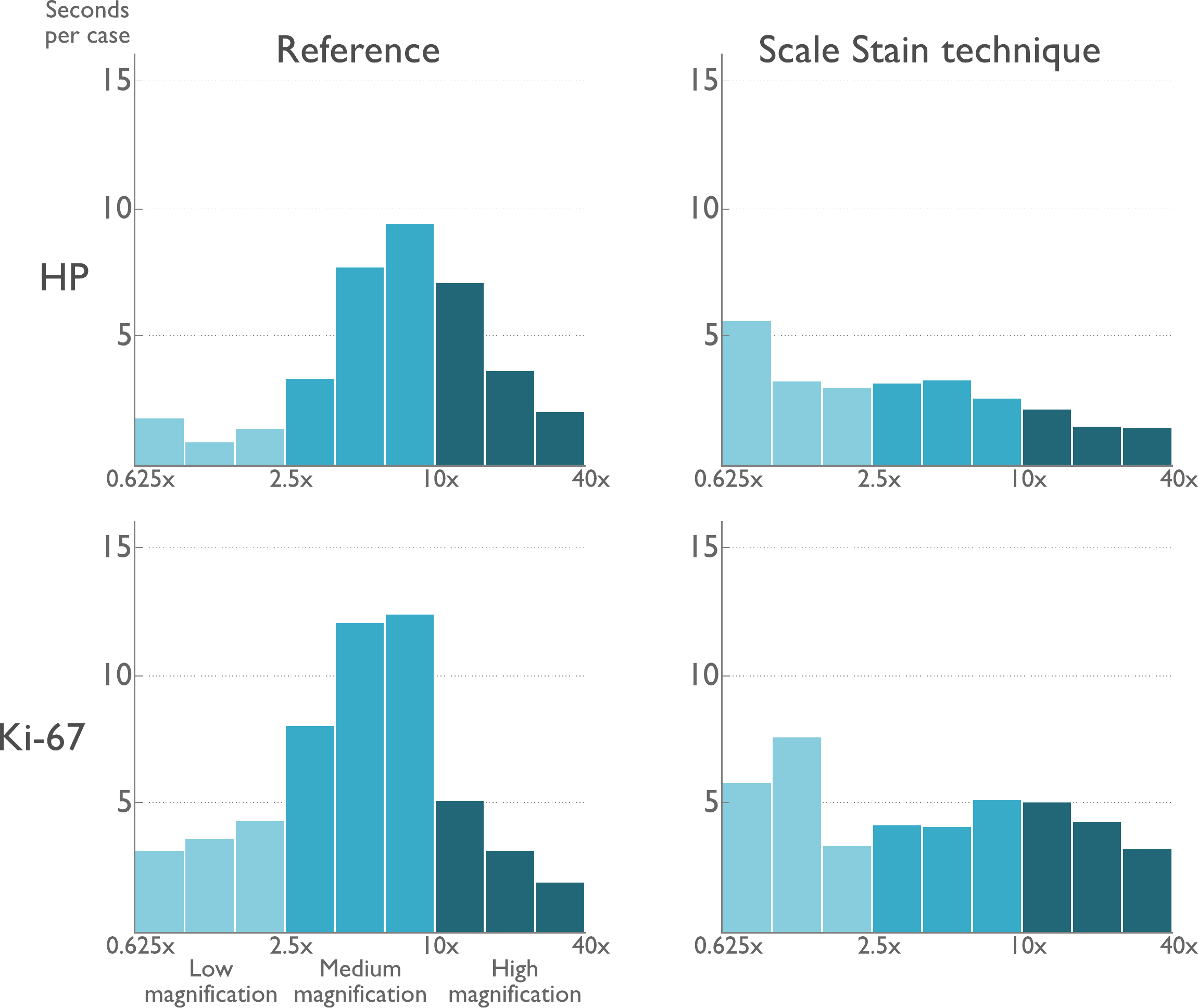}
\caption{Histogram of how many seconds the participants spent on each zoom-level for both techniques and both tasks on average for each slide. The participants spent more time at low magnification when the tool was used, than without it.}
\label{fig:zoomLevels}
\end{figure}

The effect of technique on strategy can also be seen by studying the zoom-level histogram in Figure \ref{fig:zoomLevels}. More time was spent in the medium magnification in the reference condition, whereas the magnification levels were evenly spread out with the Scale Stain technique. Similarly to the other results, the task had little effect on the magnification level used to solve the problem at hand.

\begin{figure*}
\centering
\includegraphics[width=\textwidth]{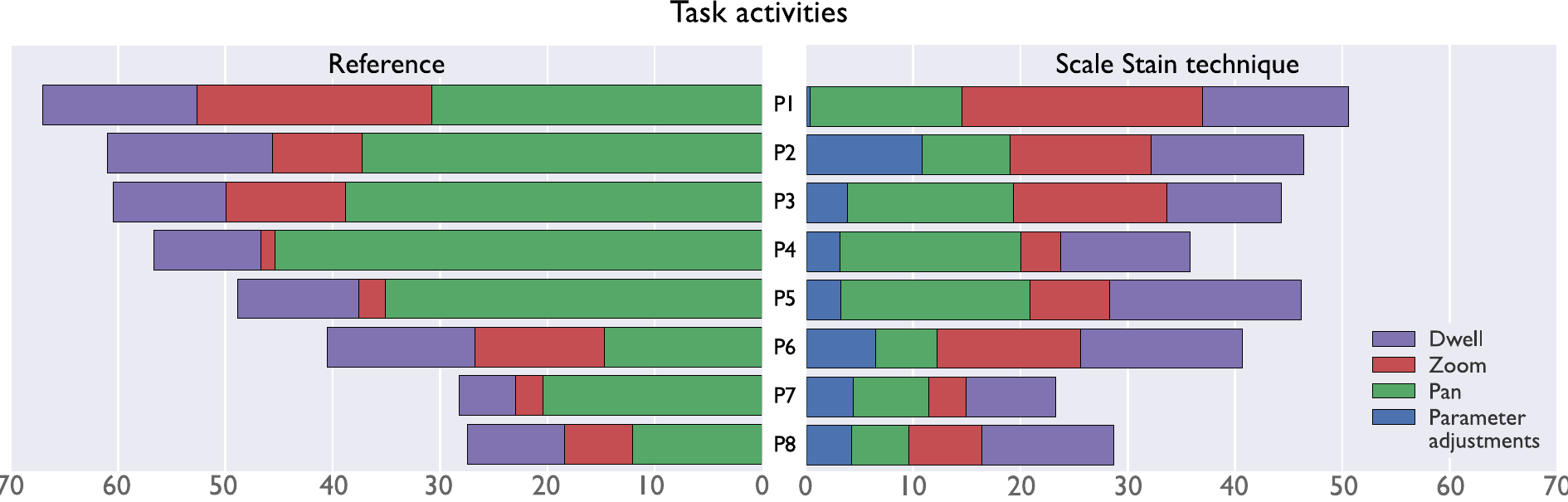}
\caption{The percentage of different activities: Contrast searching, panning, zooming and dwell per participant and trial. The left plot is the behavior when no tool was used, whereas the right plot is when the tool was used.}
\label{fig:taskActivities}
\end{figure*}

The difference in exploration strategy can also be seen as the time spent on different tasks. In Figure \ref{fig:taskActivities}, the amount of panning, zooming, dwell and parameter adjustments has been measured as the dominating activity within each second for all trials, comparing different participants for both techniques. The activities for both tasks have been merged, since all participants had the same behavior for both tasks. In the Scale Stain condition, most participants spent some time performing parameter adjustments, which adds to the total time. On the other hand all participants perform considerably less panning, which was the main reason for the $15\%$ increase in task efficiency.

\subsection{User perception}
The visualization tool was well received by all participants, who could all imagine using it in clinical practice. The tool was perceived as improving the overview of the whole slide, with less risk of missing an important area as a consequence. In contrast, the tasks performed without the tool were perceived as being tedious. Two participants even stated that they gave up the search for a hotspot and just took something due to the time consuming search needed in the reference condition.

The users also commented on different tool independent error sources: One of the main difficulties was the presence of false positives. In the Ki-67 task there were both areas with positively stained lymphocytes and in-situ component, which should be ignored. In the Helicobacter task, staining components were detected in areas where bacteria impossibly could survive, or could not be confirmed morphologically by inspecting the stained components in the highest magnification.

Three additional difficulties with the Ki-67 task were mentioned. First, the lack of a clear hotspot made it hard to decide on what area to choose. Second, there was sometimes a mismatch between the detected size of the hotpot and the size of the circle that was used to mark the hotspot. One user expressed an urge to buckle the shape of the circle a little bit to be able to fit all the positive cells within it. Another user explained the choice of the hotspot as a two-step process: In the first step, you zoomed in the most active area and in the second step the actual area to be counted was selected by placing the circle within the visible area. Two related difficulties were also mentioned. In the reference condition, it was difficult to visually remember how much staining there was in different areas, and in both conditions, it was hard to separate high density of positive cells from high staining intensity. Three pathologists dealt with these two problems by changing their cognitive strategy: Instead of trying to remember different staining intensities, the number of cells in different areas were quickly counted and only the area with the highest number were remembered.

The parameter space picker was perceived as being easy to use and all pathologists understood approximately the dimensions of the triangle. As could also be detected in the behavioral traces in the previous section, three participants used the triangle quite differently. \textit{P1} stated that when going from Helicobacter task to the Ki-67, you had to lower the sensitivity. \textit{P2} who used the picker tool the most, described the strategy as going slowly from a low to high sensitivity until the first hotspot popped out in the image, and used the pop out effect as the hotspot selection criteria. P6 assumed that a high sensitivity should be preferred, but sometimes increased the tumor visibility for cases where the tumor couldn\'t clearly be detected.

During the experiment, two noteworthy special strategies were observed. First, an interesting decision was made by \textit{P8} who was one of the most efficient participants. A group of 4-5 areas was detected as possible false positives in the Scale Stain condition. By zooming in on one of them and finding out it was false, the pathologist then concluded that all of the areas were false positives without looking at the others. The participant explained this behavior by saying that it could clearly be seen that the group of areas were not true areas, and by inspecting the most uncertain area, the others could also be excluded.

\textit{P4} scanned through the whole slide even in the Scale Stain condition, this behavior was explained by the concern that the probability of finding something unexpected otherwise would decrease. This risk was however not considered a major concern, since for real cases these findings would be discovered in the mandatory H\&E stained slide.

\section{Discussion}
The use of the Scale Stain technique increased the efficiency with 15\% with maintained accuracy for two typical tasks. Multiple findings point towards this being a low estimate of the efficiency gain. First, the participants only had a very limited amount of training to learn using the tool compared to the reference condition where they performed a task that was familiar to them. Moreover, in the reference condition, two participants stated that they prematurely stopped the search because it was too tedious.

The participants were informed that the duration for each task was recorded but that it was more important to make a correct decision than performing the review quickly. This means that they probably used their gut-feeling to stop whenever they felt being in full control. The Scale Stain technique makes a good job at giving the participant that sense of control, which is probably an important reason behind the efficiency gain. This approach is in sharp contrast against earlier approaches for the Ki-67 task \cite{Niazi2016}, that automatically detect the hotspot, circle it and visualize the result as a heatmap in order to communicate as much of the algorithm’s uncertainty as possible. Here the pathologist is left of figuring out the connection between the automatic decision and the underlying image by themselves, not fulfilling R3. On the other hand, the Scale Stain technique did not improve the accuracy, why these two approaches perhaps could be combined. The algorithm can suggest a hotspot selection, and the Scale Stain visualization can be used to check whether the algorithmic choice is reasonable.

An important notice about the efficiency measurements for each technique, is that they are the effect of two quite different exploration strategies. This means that the $15\%$ efficiency gain is probably not particularly stable with changing conditions. For example, with a doubled tissue area, the amount of panning time would double in the reference condition but only add only a few extra zoom dips the Scale Stain condition. This fact makes the Scale Stain technique even more suited for larger specimens. On the other hand, if the staining density increases until the level where it becomes visible at low magnification, there is less need for the enhancement. That is,  the Scale Stain technique is most suitable for diagnostic tasks in large specimens with a sparsely distributed staining component.

The Scale Stain technique only had a modest effect on the accuracy. This is surprising considering that performing the tasks in the reference condition should be cognitively complicated. For the helicobacter task, it is easy to miss an area when scanning around in the sample. Indeed, the users often missed part of the slide, as can be seen in Figure \ref{fig:navigationTracks}(a), but these misses were rare enough that they did not have any significant effect in this small study. In the Ki-67 task, the pathologists must rely on visual working memory to perform the comparison of the overall intensity, which are too complicated to remember due to the limited size of the visual working memory. In controlled studies for more artificial tasks \cite{Plumlee2006}, this limitation has been shown to reduce the task accuracy. An important difference with this study compared to earlier studies of multi-scale systems is that pathologists are experts at solving problems by panning and zooming. This means that they have developed mental strategies to overcome inherent limitations of problem-solving in multi-scale information spaces. For the hotspot selection task, the pathologists struggled with visual memory limitations but they also mentioned using a counting strategy to overcome that limitation.

The visualization pipeline that was presented fulfilled our three requirements \textit{R1-R3}, within a specific problem domain. The problem domain consisted of finding sparse amount of staining with a known color to support the user to conclude presence and absence, or to compare the quantity of different regions. However, these requirements could also be interpreted as design guidelines when building visualization pipelines for other problems within pathology imaging or other similar problem domains. For example, it could be possible to extend the approach to extracting small edge structures, which could highlight stromal structures within tumors.

The 100\% sensitivity requirement \textit{(R1)} is not a new idea and is commonly proposed within medical image processing to automatically exclude irrelevant areas and let the user go through the remaining areas to check for false positives. It is however not always possible to reach 100\% sensitivity. If it is enough to retain the accuracy compared to the manual task, it is sufficient for the extraction algorithm to have the same sensitivity as the pathologist when manually scanning through a slide in high magnification.

The large difference between our approach and an automatic approach lies instead within \textit{R2} and \textit{R3}. By providing a relatively simple and intuitive mapping between the high and low magnification image, each zooming action becomes an opportunity to learn how the mapping works. This should result in a situation where the pathologist’s skill using the system is allowed to improve with experience. During the short duration of study, this ability was not allowed to develop for most participants, however one participant mentioned extrapolating information gained when zooming in to other areas not inspected in high magnification.

The Scale Stain system goes beyond the capability of a conventional microscope where the lens system only creates Gaussian low-magnification representations of the pathology slide. Whereas the microscope only allows you to get an overview, to zoom and get details on demand, the Scale Stain system completes the information seeking mantra\cite{Schneiderman1996} by adding filter capabilities to the review of pathology slides.

Still, the presented filter can only filter on a specific color and does currently not work for the majority of pathology slides, which are stained with H\&E. Pathology visualization is  a novel field, which needs further investigation in order to reach the same level as maturity as within volume visualization. Recent medical studies have used image processing algorithms in a controlled setting to derive statistical image features that are novel predictors of patient survival including novel stromal features \cite{Beck2011} and heterogeneity in the Ki-67 expression throughout the whole slide \cite{Laurinavicius2016}. These novel features are not easy to distinguish in the microscope, which is probably why they have not been discovered without computational aids. However, these visual patterns could be made visible with (or modifications of) the filtering approach presented in this paper. The Scale Stain technique could therefore provide a way for pathologists to double check computational results or even to discover novel morphological patterns that are not possible to see in the microscope today.

\section{Conclusions}
We have presented a novel visualization approach that brings the idea of alternative projections or filters to pathology images. This approach was enabled by pre-processing relevant visualization settings in a flexible approach that would be easy to deploy in clinical routine. The approach was implemented in a fully functional prototype that supported real-time rendering. The prototype was then evaluated in a user study where it was concluded that the pathologists used the tool to reduce the amount of tedious panning needed to perform two common clinical tasks. By using the tool the task completion time was reduced with 15\%, at maintained accuracy.

This study represents one of the first approaches for visualization of digital pathology images that go beyond reproducing glass-slide review behavior, by adding interactivity to the visualization pipeline beyond brightness, contrast and change of focus. Our work is also one of the first user studies to provide empirical evidence of increased efficiency made possible by digital tools in pathology, for routine diagnostic tasks.

\ifCLASSOPTIONcompsoc
  \section*{Acknowledgments}
\else
  \section*{Acknowledgment}
\fi

The authors wish to thank all the pathologists at Karlstad Central Hospital and Link\"{o}ping University Hospital who took the time to participate in this study. We would also like to thank the pathology engineering team at Sectra who supported the implemention of the prototype. This work was supported by VINNOVA (2013-03906) and the Swedish Research Council (2011-4138).




\bibliographystyle{IEEEtran}

\bibliography{template}

\vspace*{-2\baselineskip}

\begin{IEEEbiography}[{\includegraphics[width=1in,height=1.25in,clip,keepaspectratio]{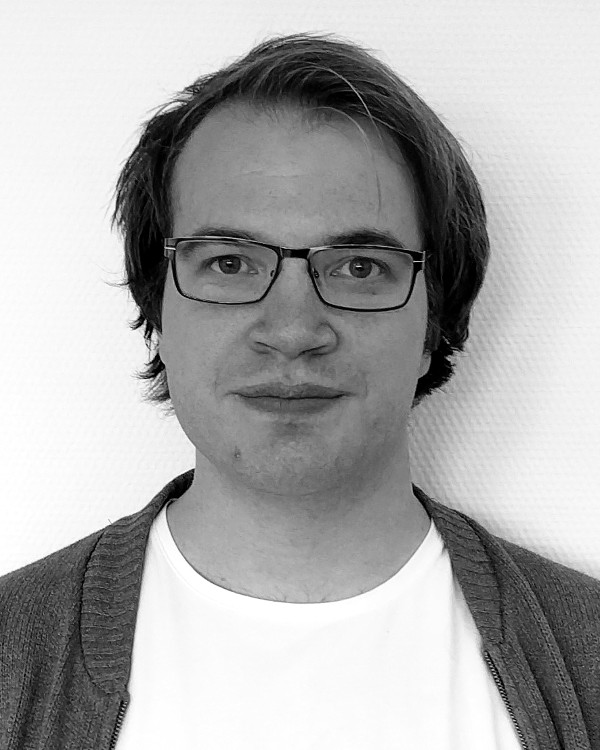}}]{Jesper Molin}
 is a PhD student in Human Computer Interaction at Chalmers University of Technology, and works as a developer at Sectra AB. He holds a MSc in Applied physics and in Biomedical Engineering from Link\"{o}ping University. His current research focus is in Human Centered Design within digital pathology, working with visualization, digital image analysis and interaction design.
\end{IEEEbiography}

\vspace*{-2\baselineskip}

\begin{IEEEbiography}[{\includegraphics[width=1in,height=1.25in,clip,keepaspectratio]{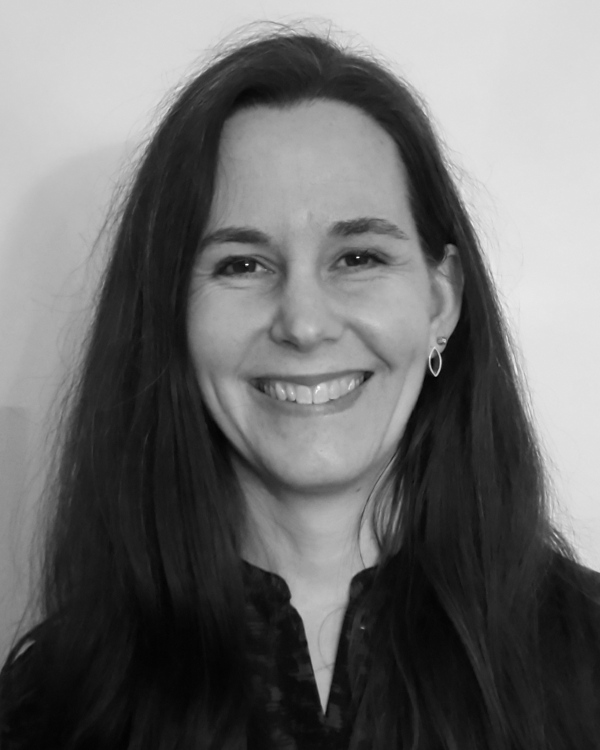}}]{Anna Bod\'{e}n}
is a clinical pathologist at Link\"{o}ping pathology department since 2010 and has been a PhD student since 2015. She is implementing and practicing digital pathology at the department. Her interests are workflow coupled to digital pathology and the possible different visualization aspects of digital pathology. Her main field is breast cancer.
\end{IEEEbiography}

\vspace*{-2\baselineskip}

\begin{IEEEbiography}[{\includegraphics[width=1in,height=1.25in,clip,keepaspectratio]{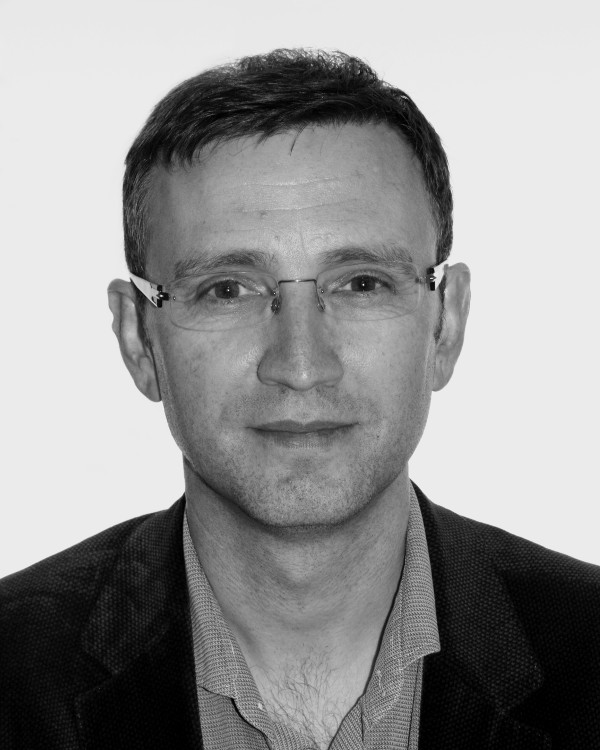}}]{Darren Treanor}
is a consultant pathologist at Leeds Teaching Hospitals NHS Trust, honorary clinical associate professor at the University of Leeds, United Kingdom, and guest professor in digital pathology at Link\"{o}ping University, Sweden. He runs the Leeds virtual pathology project, which has been carrying out digital pathology research and development since 2003. He has co-authored over 60 papers in the medical and computing literature, mostly within digital pathology and preclinical research.
\end{IEEEbiography}

\vspace*{-2\baselineskip}

\begin{IEEEbiography}[{\includegraphics[width=1in,height=1.25in,clip,keepaspectratio]{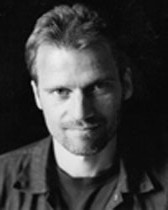}}]{Morten Fjeld}%
's research activities are situated in the field of Human-Computer Interaction with a focus on tangible, tabletop, and cross-device interaction. In 2005, he founded the t2i Interaction Lab at Chalmers. He holds a dual MSc degree in applied mathematics from NTNU (Norway) and ENSIMAG (France), and a PhD from ETH-Z (Switzerland). In 2002, Morten Fjeld received the ETH Medal for his PhD titled "Designing for Tangible Interaction". In 2011, he was a visiting professor at NUS Singapore; in 2016 he was a visiting professor at Tohoku University, Japan.
\end{IEEEbiography}

\vspace*{-2\baselineskip}

\begin{IEEEbiography}[{\includegraphics[width=1in,height=1.25in,clip,keepaspectratio]{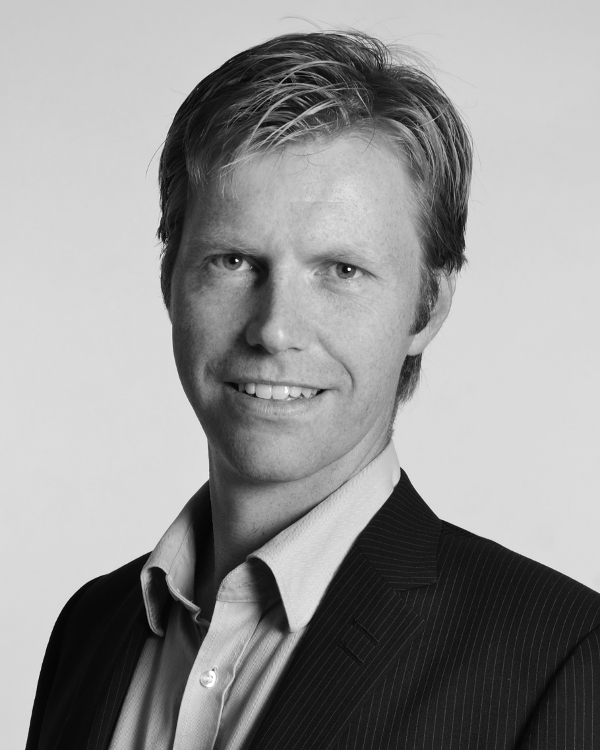}}]{Claes Lundstr\"{o}m}
 currently holds two positions, in industry as Research Director at Sectra AB and in academia as Adjunct Associate Professor at Linköping University. His primary research focus is visualization methods to enable new levels of accuracy and efficiency within medical imaging, in demanding clinical settings. A particular emphasis is given to cross-disciplinary work, considering aspects of human-computer interaction, informatics, and applied image analysis.
\end{IEEEbiography}
\end{document}